\documentclass{cmmse2018}
\usepackage{amssymb}
\usepackage[pdftex]{graphicx}
\usepackage{epstopdf}
\usepackage{xcolor}
\usepackage{amsmath}
\usepackage{hyperref}
\usepackage[utf8]{inputenc}

\begin{document}

%\pagestyle{empty}

%  Include here your own macros.

\newcommand{\Eqref}[1]{(\ref{#1})}

%  Data for the headings. Please fill both fields.

\markboth
  {A new definition of the distortion matrix} % abridged title
  {A. J. Mu\~noz-Montoro et al.}

\title{A new definition of the distortion matrix for an audio-to-score alignment system}

%\author{Full name}{e-mail}{affiliation number (below)}

\author{A. J. Mu\~noz-Montoro}{jmontoro@ujaen.es}{1}

\author{P. Vera-Candeas}{pvera@ujaen.es}{1}

\author{D. Suarez-Dou}{davidsdou@gmail.com}{2}

\author{R. Cortina}{raquel@uniovi.es}{2}

%\affiliation{number}{department}{university}

\affiliation{1}{Department of Telecommunication Engineering}{University of Ja\'en}

\affiliation{2}{Department of Computer Science}{University of Oviedo}

%  Abstract, key words and MSC codes.

\begin{abstract}
In this paper we present a new definition of the distortion matrix for a score following framework based on DTW. The proposal consists of arranging the score information in a sequence of note combinations and learning a spectral pattern for each combination using instrument models. Then, the distortion matrix is computed using these spectral patterns and a novel decomposition of the input signal.

\end{abstract}

\section*{\centering Extended abstract}
\section*{Introduction}
Audio-to-score alignment is the task of synchronizing an audio recording of a musical piece with its corresponding symbolic score. An audio-to-score alignment system, also denoted score follower, analyzes the content of the input audio at a given time point and maps it to a corresponding time point on the score.

Traditionally, the alignment process is performed in two stages: feature extraction and alignment. In the first stage, spectral features are extracted from the audio signal for characterizing the musical content of the performance, meanwhile the alignment process is focused on finding the best match between the feature sequence and the score. Classical systems rely on cost measures between score events and performance times to find this match, generating a distortion matrix which can be interpreted as the cost of the matching for each event score at each performance time. Different techniques for aligning  time series or sequences have been applied in the literature, such as Hidden Markov Models (HMMs) \cite{Duan2011Soundprism:Audio,Joder2011AMatching} and Dynamic Time Warping (DTW) \cite{Orio2001AlignmentScore,Dixon2005LiveWarping,Carabias-Orti2015AnWarping,Rodriguez-Serrano2016}.

Up to date, DTW based methods have demonstrated to provide the best alignment results in the MIREX Real-time Audio to Score Alignment\footnote{The Music Information Retrieval Evaluation eXchange (MIREX) is an annual evaluation campaign for MIR algorithms. Real-time Audio-to-Score Alignment (a.k.a. Score Following) is one of the evaluation tasks. \url{http://www.music-ir.org/mirex}.} task, standing out the results presented in \cite{Rodriguez-Serrano2016}. Rodriguez et al. proposed a score following framework based on DTW in which the score information is organized in a sequence of unique occurrences of individual and concurrent notes from several instruments, denoted as score units. \cite{Rodriguez-Serrano2016} used a MIDI synthesizer to learn a spectral pattern ${b}_k(f)$ for each score unit $k$. Then, the distortion matrix ${D}(k,t)$ is computed as the distortion generated by the input signal ${x}_t(f)$ and the synthetic spectral patterns per unit as

\begin{equation}\label{eq:basic_distortion_matrix}
{D}(k,t) = \phi_{\beta} ({b}_k(f),{x}_t(f))
\end{equation}

\noindent where $\beta$ is the parameter that controls the $\beta$-divergence. However, this approach has two main inconveniences. On the one hand, the alignment results are highly sensitive to the accuracy of the MIDI synthesizer. On the other hand, when several notes are common from consecutive score units, the distortion matrix is not able to discriminate between combinations, underperforming the results.

In this work we propose a new definition of the distortion matrix for a score following framework based on DTW. Unlike \cite{Rodriguez-Serrano2016}, here, the spectral patterns for each score unit is obtained during a training process from the Real World Computing (RWC) Musical Instrument Sound Database \cite{Goto2002RWCDatabases.,Goto2004DevelopmentDatabase}. Moreover, a novel signal decomposition for the input signal in function of each two consecutive score units is incorporated for mitigating the misalignment obtained by those cases with common notes in consecutive combinations.

\section*{Training stage}
In this stage, the score information about the notes and instruments played in the musical piece is arranged in a sequence of combinations, a.k.a score units. In order to learn all the spectral patterns, we propose to synthesize independently each combination $k$ and apply the following decomposition

\begin{equation}\label{eq:instrument_models1}
y^{k}(f,\tau) \approx g^k(\tau) \sum_{mj} \alpha^{k}_{mj} n_{mj}^k(f)   
\end{equation}

\noindent where $y^k(f,\tau)$ is the spectrum of the synthesized file, $\tau$ refers to the synthesized time, $g^k(\tau)$ is the temporal activity vector, $n_{m,j}^k(f)$ is the spectrum of each midi note $m$ and instrument $j$ which appears in the combination $k$, and $\alpha^{k}_{mj}$ represents the amplitude contribution of each
note and instrument which conform the combination $k$. Here, the spectrum of each note $n_{m,j}(f)$ is learned from isolated notes of real solo instrument recordings of the RWC database. Consequently, the basis functions of the combinations $b_k(f)$ are generated by instrument models as

\begin{equation}
     b_k(f) = \sum_{m,j} \alpha^{k}_{mj} n_{mj}^k(f)
\end{equation}

Note that, in a musical composition, only a few notes and instruments will be usually active at each combination $k$. However, it is frequent that some notes are present in consecutive combinations.

\section*{Distortion matrix computation}
For the computation of the new distortion matrix, we propose the following decomposition of each input signal frame in function of each two consecutive combination:

\begin{equation}\label{eq:instrument_models2}
x_{t}(f) = \left( \sum_{mj} a^{(k,k+1)}_{mj} n_{mj}(f) \right)_t + r_t(f),
\end{equation}

\noindent where $x_{t}(f)$ is the spectrum of the input signal at the frame $t$, $n_{m,j}(f)$ is the spectrum of the midi note $m$ and instrument $j$, $a^{(k,k+1)}_{mj}$ represents the amplitude contribution of each note and instrument for the combinations $k$ and $k+1$, and $r_t(f)$ is the residue of the approximation. Here, in order to make the procedure independent from the norm, both the input signal and the spectrum of each note are normalized as follows:

\begin{equation} \label{eq.Escale}
n_{mj}(f) = \dfrac{n_{mj}(f)}{\sqrt{\sum_{f}{n}_{mj}^2(f)}}  \quad ,  \quad x_{t}(f) = \dfrac{x_{t}(f)}{\sqrt{\sum_{f}x_{t}^2(f)}}
\end{equation}

Due to the uncorrelation between all the notes, we can consider $<\sum_{mj} n_{mj}(f)>$ as a space of dimension $n$, being $n$ the total number of notes that appear in the score. In fact, for each score units $b_k(f)$, the notes $n_{mj}^k(f)$, which conform the combination $k$, generate a subspace of dimension $n'$ ($n'\ll n$) and $\alpha_{mj}^k$ is an element of this subspace.

In this way, the distortion matrix is obtained by the computation of the Euclidean distance between the elements $a_{mj}^{(k,k+1)}$ and $\alpha_{mj}^k$ as follow

\begin{equation}\label{distortion}
    D(k,t) = \left(\sqrt{\sum_{mj}(a_{mj}^{(k,k+1)}-\alpha_{mj}^k)^2}\right)_t+||r_t(f)||^2
\end{equation}

\section*{Acknowledgements}
This work has been supported by the program \underline{SME instrument Phase II - H2020} under the project title \textit{``Beatik - Collaborative Digital Scores Platform for Classical Music''} with the project number reference \textbf{822897}.

\bibliographystyle{acm}
\bibliography{references}

\end{document}